\documentclass[prb,twocolumn,aps,epsf,showpacs,superscriptaddress,longbibliography]{revtex4-1}
\usepackage[pdftex]{graphicx}

\usepackage{dcolumn}
\usepackage{bm}
\usepackage{epsfig}
\usepackage{latexsym} 
\usepackage{amsmath}
\usepackage{amssymb}
\usepackage{array}
\usepackage{bbm}
\usepackage[colorlinks=true,citecolor=blue,linkcolor=blue]{hyperref}
\usepackage[usenames,dvipsnames]{color}			
\usepackage{array}
\usepackage{cancel}
\usepackage{ulem}

\usepackage{booktabs}
\newcolumntype{C}[1]{>{\centering\arraybackslash}m{#1}}
\AtBeginDocument{
\heavyrulewidth=.08em
\lightrulewidth=.05em
\cmidrulewidth=.03em
\belowrulesep=.65ex
\belowbottomsep=0pt
\aboverulesep=.4ex
\abovetopsep=0pt
\cmidrulesep=\doublerulesep
\cmidrulekern=.5em
\defaultaddspace=.5em
}
\usepackage{booktabs}
\newcolumntype{R}[1]{>{\raggedleft\arraybackslash}p{#1}}
\AtBeginDocument{
\heavyrulewidth=.08em
\lightrulewidth=.05em
\cmidrulewidth=.03em
\belowrulesep=.65ex
\belowbottomsep=0pt
\aboverulesep=.4ex
\abovetopsep=0pt
\cmidrulesep=\doublerulesep
\cmidrulekern=.5em
\defaultaddspace=.5em
}

\renewcommand{\>}{\rangle}
\renewcommand{\(}{\left(}
\renewcommand{\)}{\right)}

\renewcommand{\]}{\right]}

\renewcommand{\d}{\partial}

\newcommand{\eps}{\epsilon}

\newcommand{\Z}{\mathbb{Z}}
\newcommand{\T}{\mathcal{T}}

\newcommand{\Pover}{\underset{P}{\curvearrowleft}}

\begin{document}
\title{Radical chiral Floquet phases in a periodically driven Kitaev model and beyond}

\author{Hoi Chun Po}
\affiliation{Department of Physics, University of California, Berkeley, CA 94720, USA}
\affiliation{Department of Physics, Harvard University, Cambridge MA 02138, USA}

\author{Lukasz Fidkowski}
\affiliation{Department of Physics and Astronomy, Stony Brook University, Stony Brook, NY 11794, USA}
\affiliation{Kavli Institute for Theoretical Physics, University of California, Santa Barbara, CA 93106, USA}

\author{Ashvin Vishwanath}
\affiliation{Department of Physics, University of California, Berkeley, CA 94720, USA}
\affiliation{Department of Physics, Harvard University, Cambridge MA 02138, USA}

\author{Andrew C. Potter}
\affiliation{Department of Physics, University of Texas at Austin, Austin, TX 78712, USA}

\begin{abstract}
 We theoretically discover a family of non-equilibrium fractional topological phases in which time-periodic driving of a 2D system produces excitations with fractional statistics, and produces chiral quantum channels that propagate a quantized fractional number of qubits along the sample edge during each driving period. These phases share some common features with fractional quantum Hall states, but are sharply distinct dynamical phenomena. Unlike the integer-valued invariant characterizing the equilibrium quantum Hall conductance, these phases are characterized by a dynamical topological invariant that is a square root of a rational number, inspiring the label: radical chiral Floquet phases. We construct solvable models of driven and interacting spin systems with these properties, and identify an unusual bulk-boundary correspondence between the chiral edge dynamics and bulk ``anyon time-crystal" order characterized by dynamical transmutation of electric-charge into magnetic-flux excitations in the bulk.
\end{abstract}

\maketitle

\section{Introduction}
Time periodic driving serves not only as a powerful tool to engineer effective Hamiltonians\cite{oka2009photovoltaic,lindner2011floquet,wang2013observation}, but also as a means to produce intrinsically dynamical topological phases that do not exist in the static limit\cite{kitagawa2010topological,jiang2011majorana,rudner2013anomalous,von2016phaseI,else2016classification,potter2016topological,roy2016abelian,po2016chiral,roy2016periodic,harper2016stability}. Namely, by subjecting a system to a local time-dependent Hamiltonian $H(t) = H(t+T)$ with period $T$, one can realize anomalous edge dynamics that cannot be implemented by any local Hamiltonian acting only near the edge. This opens the door to new methods for coherently manipulating quantum dynamics that would otherwise be impossible in a lower-dimensional system.

Striking examples include rational chiral Floquet (CF) phases\cite{rudner2013anomalous,po2016chiral,harper2016stability} whose edges form chiral quantum channels that unidirectionally pump discrete packets of quantum information during each drive period. These phases arise in 2D systems whose bulk dynamics are trivial $U(T)_\text{bulk}\approx 1$, so that all of the action of $U(T)$ occurs in a quasi-1D strip around the sample edge. Such locality-preserving 1D time-evolution operators are exhaustively characterized by a topological invariant $\nu = \log r$, where $r$ is a rational fraction that measures the ratio of quantum information being transferred to the right vs.\ that to the left across any point in the system boundary\cite{gross2012index}. In a purely 1D system (i.e. one that is not the edge of a 2D system), one can always consider open boundary conditions, in which case there must be an exact balance of quantum state flow, $r=1$, so that states cannot pile up or be depleted from the ends of the system (otherwise the quantum dynamics cannot respect both unitarity and locality). However, the boundary of a 2D system forms a closed loop, which allows it to evade this restriction and realize any rational index, $\nu = \log r$\cite{po2016chiral}. Such rational CF phases are, loosely speaking, dynamical analogs of integer quantum Hall phases familiar from thermal equilibrium settings -- which can also occur in non-interacting systems, have ordinary bulk properties, and chirally propagating edges that are protected even in the absence of any symmetry. Despite these similarities, rational CF phases are sharply distinct from such equilibrium phenomena. For example, their edge states exhibit a discrete pumping of quantum information rather than continuous flow of heat and charge, and they have topological invariants with a completely different structure (rational vs integer).

In equilibrium settings, strong interactions can effectively fracture the original microscopic particles into emergent excitations with fractional (anyonic) statistics, leading to new types of topological behavior like the fractional quantum Hall effect. Given the rough parallels between rational CF phases and the integer quantum Hall effect, it is natural to ask: Can strong interactions also produce new ``fractional" CF phases? 

In this paper, we explore CF phases in systems in which strong interactions lead to nontrivial bulk dynamics characterized by emergent anyon excitations with fractional statistics (Abelian topological order). The presence of emergent bulk anyons with fractional statistics leading to distinct topological bulk and edge characteristics from the CF phases of unfractionalized bosons and fermions described in Ref.~\onlinecite{po2016chiral}.
Namely, the external driving can supply the energy to pump otherwise immobile or confined defects around the boundary of the system. The defects of Abelian topologically ordered systems can be non-Abelian objects with irrational quantum dimension (sometimes called twist defects or genons~\cite{barkeshli2013twist}). These non-Abelian defects rely on the presence of topological order, and arise despite the absence of mobile or deconfined non-Abelian particles. We show that the Floquet drive can induce a chiral motion of non-Abelian twist defects along the boundary, resulting in the one-way transfer of irrational amounts of quantum information along the edge during each drive period. This enables new CF phases with chiral indices that are square roots of rational numbers, inspiring the label: ``radical CF phases.'' We demonstrate an unexpected bulk-boundary correspondence between the radical CF edge and bulk dynamics that exchanges electric and magnetic anyon excitations during each period. We construct solvable, stroboscopically driven versions of Kitaev's honeycomb spin model that realize these radical CF phases, and describe how to stabilize them against drive-induced heating by fast driving\cite{abanin2015exponentially,abanin2015effective,else2016pre} or disorder induced many-body localization (MBL)\cite{abanin2015exponentially,abanin2015effective}. 

\section{Model}
We begin by constructing a solvable lattice model, which will enable controlled insight into the general structure of radical CF phases. Starting from an ordinary lattice of spin-1/2 degrees of freedom with two states per site, our strategy will be to dynamically induce $\Z_2$ topological order and subsequently liberate the emergent fermionic excitations. The edge of the model will then act as a chiral edge pump for non-Abelian (Majorana) defects of this topological order that each carry an irrational amount, $\log \sqrt{2}$, of quantum information.

We will first construct an idealized fixed-point drive with uniform couplings. 
Weakly perturbing away from this solvable point in this uniform model, leads to a long-lived pre-thermal phase\cite{abanin2015exponentially,abanin2015effective,else2016pre}, which eventually heats to an incoherent high-temperature state after an exponentially long time. This heating can be avoided 
(perhaps entirely
\footnote{Whether stable MBL can occur in dimension higher than one remains remains an important, unsettled matter of principle\cite{de2016stability}. For strong disorder, the dynamics will behave as in an MBL system at worst up to super-exponentially long timescale, and possibly forever.})
by coupling the system to a cooling bath\cite{else2016pre}, or, more strikingly, by introducing strong disorder to drive the system into a many-body localized (MBL) phase\cite{nandkishore2015many,abanin2016theory,lazarides2015fate} that remains quantum coherent without the need for cooling, as we will discuss in Sec.\ \ref{timecrystal}.

\begin{figure}[tb]
\begin{center}
\includegraphics[width=0.9\linewidth]{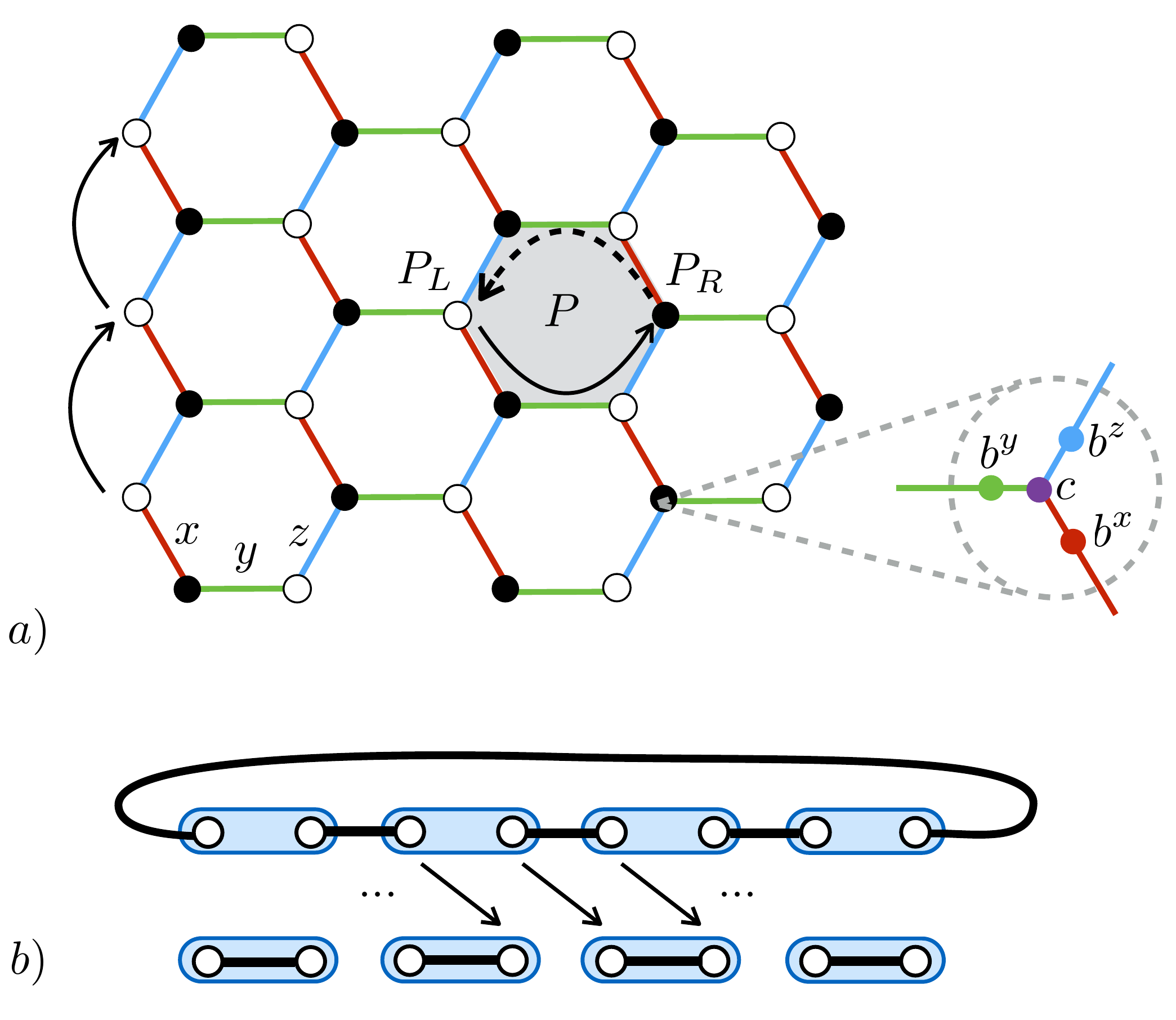}
\vspace{-.1in}
\end{center}
\caption{
{\bf Spin-1/2 Honeycomb model -- } 
(a) Depiction of a solvable lattice model for the radical chiral Floquet phase and its associated Majorana fermion description. (b) The topological and trivial phases of a chain of complex fermions (blue boxes indicate fermion sites) can be viewed as two topologically distinct ways to pair (black lines) adjacent Majorana fermions (open circles).
Under an open boundary condition, these two phase are distinguished by the presence or absence of unpaired edge Majorana fermions, but with periodic boundary condition they are related by a chiral translation of the Majorana fermions (arrows).
}
\label{fig:HoneycombModel}
\end{figure}

Our construction is based on a stroboscopically driven version of Kitaev's honeycomb spin model\cite{kitaev2006anyons}, consisting of spin-1/2 degrees of freedom, $\vec{S}_r$, sitting on sites $r$ of a honeycomb. We label the three distinct types of bonds of the honeycomb as $x$, $y$, and $z$ (Fig.~\ref{fig:HoneycombModel}a). The system is then subjected to a three-step stroboscopic time evolution obtained by sequentially applying the Hamiltonians:
\begin{align} 
H_j = \begin{cases} 
\frac{3}{T} h^{\[x\]}, & 0\leq t< \frac T3 \\ 
\frac{3}{T} h^{\[y\]}, & \frac T 3\leq t< \frac{2T}3 \\ 
\frac{3}{T} h^{\[z\]}, & \frac{2T}3\leq t< T \\ 
\end{cases};
\hspace{0.2in} h^{[j]} = \frac{\pi J}{4}\sum_{\<rr'\>\in j}S^j_rS^j_{r'},
\end{align}
where $j\in \{x,y,z\}$.
Various proposals for physically implementing such interactions in systems of cold polar molecules have been previously presented~\cite{micheli2006toolbox,gorshkov2013kitaev}. However, for our purposes, this model serves simply as a tractable platform to theoretically explore the novel phenomena of chiral Floquet phases in fractionalized systems.

The resulting time evolution for one period is:
\begin{align}
U(T) = \T e^{-i\int_0^T H(t)dt} = e^{-i h^{[z]}}e^{-i h^{[y]}}e^{-i h^{[x]}}
\label{eq:drive}
\end{align}
where $\T$ denotes time-ordering. In the limit of weak driving ($J\ll 1$), $U(T)$ realizes a conventional static phase with $\Z_2$ topological order featuring an emergent gapless Majorana fermion\cite{kitaev2006anyons}. However, we will instead consider the strong driving limit with $J = 1$.

Following Ref.\ \onlinecite{kitaev2006anyons}, this model can be solved by writing each spin-1/2 in terms of four Majorana fermion variables, $\{c_r,b^{x,y,z}_r\}$, as:
\begin{align}
S^j_r = ic_r b^j_r ,
\end{align}
This fermion description has extra artificial degrees of freedom not present in the original spin model, corresponding to a $\Z_2$ gauge redundancy generated by $(c_r,\vec{b}_r)\rightarrow (-1) (c_r,\vec{b}_r)$, and must be subjected to the gauge-neutral sector via the on-site constraints $\(-iS^x_rS^y_rS^z_r\) = \(c_rb^x_rb^y_rb^z_r\) = 1$ in order to faithfully describe the spin-1/2 system. We can draw the Majorana fermion degrees of freedom such that $c_r$ resides on the honeycomb sites, and $b^i_r$ reside on the links of type $i$ (see Fig.~\ref{fig:HoneycombModel}a). It is convenient to pair the $\vec{b}_r$ Majorana operators into $\Z_2$ gauge link variables $\sigma_{r,r'} = ib^{j}_rb^{j}_{r'}$, where $j\in \{x,y,z\}$ according to the type of link $\<r,r'\>$, and where we take an arbitrary fixed orientation of $r\rightarrow r'$ on each type of bond.

Each factor of $e^{-ih^{[j]}}$ ``hops" the $c$-Majoranas:
$e^{ih^{[j]}}c_re^{-ih^{[j]}} = c_{r+\hat{e}_j}\sigma_{r,r+\hat{e}_j}$
where $\hat{e}_j$ is the oriented unit vector along the type-$j$ bonds. The gauge link variables $\sigma_{rr'}$ are invariant under the Floquet evolution, which we can express as a conservation of gauge flux, $\mathcal{F}_P$, through each hexagonal plaquette, $P$:
\begin{align}
U(T)^\dagger \mathcal{F}_P U(T)^{\vphantom\dagger} = \mathcal{F}_P; ~~\mathcal{F}_P = \prod_{\<rr'\>\in \partial P} \sigma_{r,r'}.
\end{align}
In the bulk, the $c$ Majorana fermions are driven in small counter-clockwise loops, encircling their respective plaquettes after two driving periods (Fig.~\ref{fig:HoneycombModel}a) and accumulate a $\Z_2$-valued Aharonov-Bohm phase $\mathcal{F}_P$ along the way. While seemingly innocuous at first glance, this phase will play a crucial role in enabling the radical CF edge physics. 

We emphasize that, while the above model provides an effective description in terms of Majorana fermion degrees of freedom, it arises from a pure spin model with no microscopic fermions. Rather, the fermions and $\Z_2$ gauge fluxes are \emph{emergent} anyonic degrees of freedom that arise from the special character of the Floquet drive.

\section{Topological dynamics}
Next, we analyze the topological aspects of dynamics in this driven honeycomb spin model. In particular, we will show that the edge dynamics in one Floquet cycle corresponds to a unit translation of the emergent Majorana fermions, whereas the bulk features a form of Floquet enriched topological order amounting to a dynamical anyon transmutation between the electric and magnetic excitations.

\subsection{Edge chiral transport}
In contrast to their bulk counterparts, the $c$-Majoranas at the edge are driven in a large clockwise loop around the entire system boundary, such that a single Majorana fermion crosses any cut through the edge during each driving cycle. Since each Majorana degree of freedom has ``half" the number of degrees of freedom as a fermion, it corresponds to $\sqrt{2}$ quantum states, and hence we expect a radical chiral unitary index $\nu = \log \sqrt{2}$. 

We can confirm this expectation by the following trick: instead of computing the index $\nu$ for $U(T)$ directly, we can consider the time evolution for two periods, $U(2T)$. Since the dynamics in each period are identical, $U(2T)$ transfers twice the amount of quantum information as $U(T)$, implying $\nu_{2T} = 2\nu$. Unlike $U(T)$, however, $U(2T)$ is governed by a rational chiral unitary invariant, $\nu_{2T}$, that can be computed by the algebraic method of Refs.~\onlinecite{gross2012index,po2016chiral}. The quantum information pumped across a cut in the edge during each period is quantified by taking a basis of observables on one side of the cut, evolving them forward in time, and evaluating its overlap with observables on the other side. The chiral topological index is then the difference in quantum information being pumped to the right vs.\ that being pumped to the left per period. A direct computation (Appendix \ref{supp:index}), using strings of spin operators $S^{x,y,z}_i$ as a basis for operators, shows that  $\nu_{2T}= \log 2$, i.e.,
\begin{align} 
\nu = \frac12\nu_{2T} = \log \sqrt{2},
\end{align}
confirming that this model indeed realizes a radical CF phase.

\subsection{Bulk anyon transmutation}
To characterize the bulk dynamics, we can pair the bulk Majoranas on the left and right sides of each hexagonal plaquette into complex fermion orbitals, $\psi_P = \frac12(c_{P_L}+iW_{\Pover} c_{P_R})$, where $\displaystyle W_{\Pover}\equiv\sigma_{_{\swarrow}}  \sigma_{_\leftarrow}\sigma_{_{\nwarrow}}$ is a gauge string connecting the sites $P_{R/L}$ via a counterclockwise loop over the top of plaquette $P$ (dashed curved arrow in Fig.~\ref{fig:HoneycombModel}a). The plaquette fermion orbitals can be either occupied or empty, corresponding to local fermion parity: $\mathcal{P}_P = (-1)^{\psi^\dagger_P\psi^{\vphantom\dagger}_P} = \pm 1$ respectively. During each drive period the plaquette parity acquires an Aharonov-Bohm phase on plaquettes with gauge flux: $\mathcal{P}_P\rightarrow \mathcal{F}_P\mathcal{P}_P$. In other words, the plaquette fermion parity is conserved on gauge-flux-free plaquettes ($\mathcal{F}_P=+1$), and is flipped on plaquettes with a flux ($\mathcal{F}_P=-1$). 

To physically interpret this result, first note that the absence of quantum dynamics for the bulk gauge degrees of freedom indicates that the Floquet operator induces dynamical $\Z_2$ topological order, with three topologically distinct types of anyon excitations: a fermion $\psi$ ($\mathcal{P}_P\hspace{-3pt}=\hspace{-3pt}{-1}$, $\mathcal{F}_P\hspace{-3pt}=\hspace{-3pt}{+1}$), a bosonic flux $m$ ($\mathcal{P}_P\hspace{-3pt}=\hspace{-3pt}{+1}$, $\mathcal{F}_P\hspace{-3pt}=\hspace{-3pt}{-1}$), and their bosonic bound state: $e=m\times \psi$. Each of these types of excitations has mutual statistics $(-1)$ with the others. 
During each driving period, the number of $\psi$ excitations is conserved, but the $e$ anyons are transmuted into $m$ anyons and vice versa, a phenomena dubbed Floquet enriched topological order (FET)\cite{potter2016dynamically}. 

While we have illustrated these phenomena for a special point in the phase diagram of the model, the addition of disorder into Eq.~\ref{eq:drive} can produce many-body localization (MBL), and stabilize these properties over an extended region of of parameter space.
Since the bulk Floquet evolution does not produce kinetic motion of any of the bulk anyon particles, it appears naturally amenable to MBL. Indeed, as explained in Ref.\ \onlinecite{potter2016dynamically}, one can argue that a disordered version of this Hamiltonian produces a stable MBL phase, albeit one that exhibits discrete time-translation symmetry breaking\cite{else2016floquet,khemani2015phase,von2016phaseII,von2016absolute} due to the continual period-$2T$ flip-flopping of $e$ and $m$ particles\cite{potter2016dynamically}. We will expand on these arguments in Sec.\ \ref{timecrystal}.

%

\section{Bulk-boundary correspondence}
In the solvable point of the driven honeycomb spin model described above, we saw that the radical chiral Floquet nature of the edge was accompanied by bulk FET order. Here, we will argue that these two phenomena are always linked. 
To build some intuition, we first recall that the $e\leftrightarrow m$ exchanging FET order can be viewed as arising from a dynamical pumping of loops of 1D topological chains of the $\psi$-fermions onto the system boundary during each Floquet period\cite{potter2016dynamically}, which toggles the 1D chain of fermions at the edge between the topological and trivial phases. Since the fermion parity of the 1D topological fermion chain is flipped by insertion of a $\pi$-flux\cite{kitaev2001unpaired}, this pumping adds a fermion to each bulk gauge flux, thereby interchanging $e$ and $m$ particles. 
By formally decomposing each complex fermion degree of freedom at the edge into a pair of Majorana fermions, one can see that the chiral Majorana translation at the edge of the radical CF phase toggles the 1D topological invariant of the edge (Fig.~\ref{fig:HoneycombModel}b). 
In this picture, the chiral translation of Majorana defects at the edge by an odd number of sites modifies the pairing patterns of the Majorana fermions, which amounts to changing the topological phase of this 1D complex fermion chain\cite{kitaev2001unpaired}.

To support this heuristic picture, we will specialize to the limit of non-interacting Majorana fermions in a static gauge-flux background. In this limit, we can solve the dynamics by first specifying a gauge-flux sector in the bulk, and then evaluating the action of $U(T)$ on the Majorana fermion operators. The generalization of this argument to interacting fermion systems can be found in Ref.~\onlinecite{fidkowski2017interacting}, where we discuss the general form of Floquet bulk-edge decoupling\cite{po2016chiral} in the presence of static bulk gauge fluxes, and generalize the rigorous machinery of Ref.\ \onlinecite{gross2012index} to incorporate fermionic super-algebras.

\subsection{Chiral edge invariant}
We will begin by identifying a chiral edge invariant. 
We will consider a large finite cylinder, for which, due to the localized nature of $U$, we can consider the restriction of this evolution to a finite strip near one end of the cylinder:
\begin{align}
U^\dagger(T) c_{r}U(T)=\sum_{r'} O_{r,r'}c_{r'}.
\end{align} 
Here, $O_{r,r'}$ is an orthogonal matrix, whose indices, $r,r'$ label positions along the edge of the cylinder.

Denote the number of chiral Majorana edge modes associated with $O$ as $C$, which is related to the chiral unitary invariant by $\nu = C\log \sqrt{2}$.
With translation invariance, $C$ can be computed by the momentum space winding number via a simple generalization of the results of Ref.~\onlinecite{rudner2013anomalous} to Majorana fermions:
\begin{align} 
C \overset{\text{\tiny trans.~inv.}}{=} \int \frac{dk}{2\pi}\, \text{tr}\(\tilde{O}^{-1}(k)i\d_k \tilde{O}(k)\),
\label{eq:cleanC}
\end{align}
where $\tilde{O}_{\alpha,\beta}(k) = \int dx \, e^{ikx}O_{x,\alpha;0,\beta} $, and the trace is over the flavor indices $\alpha,\beta$.

Since we are interested in disordered systems, we would like to reformulate this invariant in a way that does not rely on  momentum conservation. A useful formal tool is to replace the integral over momentum in Eq.~\eqref{eq:cleanC} by an adiabatic flow under the insertion of ``flux". Though the fermion charge is not conserved in the present problem with Majorana fermions, we can still formally define a version of $O$ with flux $\theta\in (-\pi,\pi]$ threaded through the bond between $x=0$ and $x=1$ along the edge:
\begin{align}
(O_\theta)_{xx'} = \begin{cases}
O_{xx'}e^{i\theta} & \text{for}~-\frac{L_x}{2}<x'\leq 0<x<\frac{L_x}{2} \\
O_{xx'}e^{-i\theta} & \text{for}~-\frac{L_x}{2}<x\leq 0 <x' <\frac{L_x}{2}\\
O_{xx'} & \text{otherwise}
\end{cases},
\end{align}
where we have suppressed the flavor index, and $L_x$ denotes the circumference of the cylinder edge.

A minor, but formally necessary technical detail is that some truncation scheme is required to make the flux insertion compatible with periodic boundary-conditions. While various equivalent methods are possible, here, we have simply turned off the $e^{i\theta}$ phase twist at a distance $L_x/2$ from the origin. The effects of this finite-size truncation can be safely ignored in large systems. Namely, since $|O_{xx'}|$ results from finite time evolution with a local (2D) Hamiltonian, the spatial extent matrix elements are constrained by a Lieb-Robinson bound, which gives rise to a length scale $\ell_{\rm LR}$ set by the finite time $T$ and the (maximum) Lieb-Robinson velocity associated with the instantaneous Hamiltonians, i.e. $|O_{x,x'}|$ falls off exponentially as $\sim e^{- |x-x'|/\ell_{\rm LR}}$ for distances $|x-x'| > \ell_{\rm LR}$.
For similar reasons, $O_\theta$ will be exponentially close to a unitary matrix: $\left| \left| O^{\vphantom\dagger}_{\theta} O_{\theta}^\dagger- \mathbb {I} \right| \right| \lesssim e^{- L_x/\ell_{\rm LR}}$.

By introducing the adiabatic flow parameterized by $\theta$, the chiral edge invariant can now be written as \cite{titum2016anomalous}:
\begin{align}
C &= \int_{-\pi}^\pi \frac{d\theta}{2\pi } ~\text{tr}\(O_\theta^{\dagger} i \frac{\d}{\d\theta} O_\theta^{\vphantom\dagger}\),
\label{eq:fluxC}
\end{align}
where the trace runs over all spatial and  flavor indices.
One can verify this reproduces Eq.~\eqref{eq:cleanC} for translation-invariant edges. If $O_\theta$ were exactly unitary, then $C$ would be a winding number which is precisely quantized to integer values. For a large but finite $L_x$, this integer quantization is accurate up to exponentially small corrections of order $e^{-L_x/\ell_\text{LR}}$, due to the truncation at $x=\pm L_x/2$, and becomes exact as $L_x\rightarrow \infty$. 

We remark in passing that this invariant can also be formulated directly in the limit of an infinitely long edge, $L_x=\infty$, where the flux-threading can then be implemented by a unitary operator: $O_\theta = e^{i\theta \mathcal P} Oe^{-i\theta \mathcal P}$, where $\mathcal P$ is a projection into the subspace with $x>0$ 
\begin{align}
\mathcal P_{xx'} = \delta_{xx'} \equiv \begin{cases} 1 & \text{for}~x>0 \\ 0 & \text{for}~x\leq 0 \end{cases}.
\end{align}
Putting this form of $O_{\theta}$ into Eq.~\eqref{eq:fluxC}, one finds
\begin{align}
C = \text{tr}\( O^{-1}[\mathcal P,O]\),
\end{align}
In this form, $C$ is the trace of a difference between two projection operators, whose eigenvalues are $0$ or $1$, and therefore $C$ is precisely quantized to an integer and cannot be altered by smooth deformations (local unitary transformations of $O$). Related quantites were identified in Refs.~\onlinecite{kitaev2006anyons,gross2012index}  as a `flow' index for causal unitary matrices.

\subsection{FET bulk invariant}
To diagnose the FET order, we would like to compare the change in fermion parity with and without a $\pi$ flux threading the edge. In the FET phase, Floquet evolution toggles the edge between topological and trivial states, and hence pumps an opposite amount of fermion parity dependent on the presence or absence of a $\pi$ flux. In a non-FET phase, the parity pumped is independent of the flux. Such parity pumping is captured by comparing the determinants of $O$ with and without a $\pi$ flux inserted. To see this, observe that the fermion parity operator for the edge, $P_{F,{\rm edge}} = i^{N_\text{sites}}\prod_r c_r$, evolves as:
\begin{align}
U(T)^\dagger P_{F,{\rm edge}} U(T) =  \left( \det{O} \right) P_{F,{\rm edge}},
\end{align}
which follows from the antisymmetry of the fermion product and the orthogonality of $O$. Here, $N_\text{sites}$ is the number of Majorana sites and the phase factor is chosen to ensure $P_{F,{\rm edge}}^2=1$.

From these considerations, we can write the FET invariant as a comparison between $O_{0}$ and $O_\pi$:
\begin{align}
\mathcal{I}_\text{FET} = \det \(O_\pi O_0^{-1}\),
\label{eq:IFET}
\end{align}
which is $-1$ in the FET phase, and $+1$ otherwise. We note in passing that various equivalent forms for $\mathcal{I}_\text{FET}$ like $\(\det O_\pi /\det O_0\)$, or $\(\det O_\pi \cdot \det O_0\)$, are possible. However, the above formulation is convenient as it remains well defined in the infinite-size limit.

\subsection{Relation between $C$ and $\mathcal{I}_\text{FET}$}
From these formulations it is straightforward to relate the chiral edge and bulk FET invariants. Since $O$ is real ($O_\theta = O_{-\theta}^*$):
\begin{align}
C= \int_{-\pi}^{\pi} \frac{d\theta}{2\pi} \text{tr} \left( O_{\theta}^{\dagger} i \partial_\theta O_{\theta} \right)
= \int_{0}^{\pi} \frac{d \theta}{\pi} \text{tr} \left( O_{\theta}^{\dagger} i \partial_\theta O_{\theta} \right), 
\end{align}
and therefore
\begin{equation}\begin{split}\label{eq:}
\mathcal{I}_\text{FET} &= \det \(O_\pi O_0^{-1}\) \\
&= \exp\left( {\text{tr}\left( \log O_\pi  - \log O_0  \right)}   \right) \\
&= \exp\left(- i \pi \int_0^{\pi} \,\frac{d \theta}{\pi} \,i  \partial_{\theta} {\text{tr} \log O_\theta  }   \right)\\
&= \exp\left(- i \pi \int_0^{\pi} \,\frac{d \theta}{\pi} \,{\text{tr} \left(   O_\theta^\dagger i  \partial_{\theta}  O_{\theta}  \right) }   \right)\\
&=e^{- i \pi C}.
\end{split}\end{equation}
This establishes the bulk-edge correspondence:
\begin{align}
e^{2\pi i\nu_\text{edge}/\log2} = \mathcal{I}_\text{FET},
\label{eq:bulkedge}
\end{align}
 between the edge chiral unitary invariant and the bulk FET invariant, in the limit of vanishing gauge fluctuations and non-interacting emergent fermions. A more formal proof that also applies to the general interacting case can be obtained using super-algebra methods\cite{fidkowski2017interacting}.

The concurrent appearance of FET order also explains how this model can exhibit irrational values of the chiral edge index.  If the Floquet evolution was to factorize into commuting bulk and edge components, the chiral unitary index would necessarily be rational \cite{gross2012index,po2016chiral}. Note that the presence of bulk topological order alone is not sufficient for eluding  the rational restriction. As an example, $U(2T)$ of our model is topologically ordered and has a rational index.  However, this decomposition fails in a radical CF phase precisely due to the presence of the FET order. Specifically, in the sector with an odd number $\Z_2$ gauge fluxes in the bulk, the Floquet evolution transfers an odd number of fermions from bulk to boundary, such that the bulk and boundary factors in $U(T)$ would be anti-commuting fermionic operators. This failure to factorize exposes a loophole in the rational classification\cite{po2016chiral}, and allows for radical chiral edge invariants.

\section{Stability from strong disorder\label{timecrystal}}
Thus far, we have analyzed in detail a special zero-correlation-length point of the driven honeycomb model,  and derived a bulk-boundary correspondence which is applicable as long as the gauge fluxes are non-dynamical and the emergent Majorana fermions are noninteracting. In particular, the derived bulk-boundary correspondence is compatible with the introduction of disorder.
Next, we discuss how strong disorder can produce a bulk MBL phase and subsequently lends rigidity to the described physical properties under  the incorporation of small perturbations.

To this end, we add a fourth driving step with strongly disordered random coupling to the local conserved quantities of the clean driving steps in Eq.~1, so that $U(T)$ becomes $\tilde{U}(T)=e^{-ih_\text{dis}}e^{-ih^{[z]}}e^{-ih^{[y]}}e^{-ih^{[x]}}$, with
\begin{align}
h_\text{dis} = -\sum_P\sum_{a = e,m,\psi} \mu_{a,P} n_{a,P}.
\end{align}
Here, $\mu_{a,P}$ is a random potential for an anyon excitation of type $a$ on plaquette $P$, and $n_{a,P}$ denotes the corresponding number operator. 

The local plaquette fermion number, $n_{\psi,P}$, and the total gauge flux $n_{e,P} +n_{m,P}$ are conserved by the clean part of the drive, $U(T)$. However, the difference between the number of $e$ and $m$ particles is flipped by $U(T)$ due to the FET order. Although $h_\text{dis}$ does not fully commute with $U(T)$, we can still readily write down the exact eigenstates of the disordered drive, $\tilde{U}(T)$. 

Denote a fixed configuration of anyon excitations by $\mathcal{C}$, and let $\mathcal C'$ be the related configuration obtained by interchanging all $e$ and $m$ particles in $\mathcal{C}$. 
We can write down the energy of configurations $\mathcal{C}$ and $\mathcal{C}'$ with respect to the disorder Hamiltonian $h_\text{dis}$:
\begin{align}
E_\mathcal{C} &= -\sum_{a,P} \mu_{a,P} \, n_{a,P}(\mathcal{C}) \equiv E_0(\mathcal{C})+\Delta E(\mathcal{C});
\nonumber\\
E_{\mathcal{C}'} &= -\sum_{a,P} \mu_{a,P} \, n_{a,P}(\mathcal{C})\equiv E_0(\mathcal{C})-\Delta E(\mathcal{C});
\nonumber\\
E_0(\mathcal{C}) &= -\sum_P\mu_{\psi,P} \, n_{\psi,P}(\mathcal{C})+\frac{\mu_{e,P}+\mu_{m,P}}{2}n_{e,P}(\mathcal{C});\, \&
\nonumber\\
\Delta E &= -\sum_P\frac{\mu_{e,P}-\mu_{m,P}}{2} \, n_{e,P}(\mathcal{C}).
\end{align}

From this, we can readily identify a pair of Floquet eigenstates of $\tilde{U}(T)$
\begin{align}
|\psi_\pm\> = \frac{1}{\sqrt{2}}\(e^{i\Delta E/2} |\mathcal{C}\>\pm e^{-i\Delta E/2}|\mathcal{C}'\>\),
\label{eq:tcstate}
\end{align}
which have quasi-energies $\eps_+ = E_0(\mathcal{C})$, $\eps_- = E_0(\mathcal C)+\pi$, i.e., they differ exactly by $\pi$. Such eigenstate structure is reminiscent of that for a Floquet time crystal with spontaneous period doubling \cite{else2016floquet}.
In the present problem, a system initially prepared in state $|\mathcal{C}\>$ will oscillate between $|\mathcal{C}\>$ and $|\mathcal{C}'\>$ from one period to the next, provided that $\mathcal C$ contains at least one $e$ or $m$ excitation such that $\mathcal C \neq \mathcal C'$.

The discussion above, however, does not fully characterize the eigenstate degeneracy of the system.
This is because, for a generic configuration $\mathcal C$, there are actually an extensive number of other pairs of eigenstates that also have quasi-energies $\(E_0 (\mathcal C) ,E_0 ( \mathcal C)+\pi\)$, which can be obtained simply by flipping any subset of $e$ particles in $\mathcal{C}$ with $m$-particles, resulting in an overall degeneracy
\begin{align}
D(\mathcal{C}) = 2^{\sum_{P}\(n_{e,P}(\mathcal{C})+n_{m,P}(\mathcal{C})-1\)}
\end{align}
for each of the quasi-energies $E_0(\mathcal C)$ and $E_0(\mathcal C)+\pi$. We remark that such degeneracy implies the $e$ and $m$ particles behave effectively as entities with a quantum dimension of $2$ under our radical CF drive \cite{potter2016dynamically}.

This degeneracy will be resolved by quantum fluctuations upon moving away from the zero-correlation length limit, by even an infinitesimal amount. However, due to the FET structure, the degenerate states cannot be split in such a way that preserves both their localized properties and time-translation symmetry\cite{potter2016symmetry,potter2016dynamically}.  

At strong disorder, a natural outcome is for quantum fluctuations to spontaneously select time-crystalline MBL states of the form shown in Eq.~\eqref{eq:tcstate}. To proceed, let us consider the evolution for two periods, $U(2T)$ which can be written in the absence of an edge, as evolution under some effective topologically ordered Hamiltonian $H_\text{eff}$, with a symmetry between $e$ and $m$ particles, which derives from the dynamical permutation of $e$ and $m$ particles in $U(T)$\cite{potter2016dynamically}. Let us consider moving away from the zero-correlation-length limit by applying a generic but weak $T$-periodic perturbation, which corresponds to an $e\leftrightarrow m$ symmetry preserving perturbation to $H_\text{eff}\rightarrow H_\text{eff}+V$. Starting from a localized anyon configuration $\mathcal{C}$, we can restrict our attention to the degenerate space of anyon configurations with the same quasi-energy (modulo $\pi$) as $\mathcal{C}$, which we can model as a fermionic Hilbert space where each $e/m$ particles is a fermion site that can either be occupied or empty. The perturbation $V$ induces quantum fluctuations that mix these degenerate states, which can be viewed as virtual anyon particle-hole pair fluctuations.

To obtain a controlled description, we will assume that there is a low density of $e$ and $m$ particles in $\mathcal{C}$, with a typical separation $r$. By ``low density,'' we mean $r$ is much larger than the localization length scale $\xi \approx 1/\log(\Delta\mu_{\psi}/\Gamma_0)$, where $\Delta \mu_{\psi}$ denotes the root-mean-square variation of the disorder potential $\mu_{\psi}$, and $\Gamma_0$ denotes the strength of quantum fluctuations.


There are two distinct types of important virtual processes: First, a virtually excited fermion landing on an $e$ ($m$) particle and converts it into an $m$ ($e$) particle (such processes must occur in pairs to stay within the degenerate manifold of states associated with $\mathcal{C}$). With this process alone, we can model the system as a free fermion system, with a lattice of fermion sites corresponding to either $e$ or $m$ particles, which we will label by sites $i$, governed by the free fermion Hamiltonian $H_\psi \approx \sum_{ij}\Gamma_{ij}\psi_i\psi_j +\text{h.c.}$, where $\Gamma_{ij}\approx \Gamma_0e^{-r_{ij}/\xi}$ are generically exponentially decaying in the distance between $i$ and $j$, and $\psi_i$ destroys a fermion on site $i$.

The second type of virtual processes of interest are those in which a pair of virtually excited $e$ particles (or an equivalent pair of virtual $m$ particles) encircles a pair of fermion ``sites,'' which gives a topological phase depending on the fermion occupation numbers of the sites. This corresponds to an interaction term between fermions $H_\text{int} \approx \sum_{ij}V_{ij}\(\psi_i^\dagger\psi_i^{\vphantom\dagger}-\frac12\)\(\psi_j^\dagger\psi_j^{\vphantom\dagger}-\frac12\)+\dots$, with $V_{ij}\approx \Gamma_0 e^{-2r_{ij}/\xi}$, and where the ``$\ldots$'' indicates contributions from virtual fluctuations that encircle higher numbers of fermion sites, which are suppressed by exponential distance factors compared to the leading term.

The problem of solving for the excited eigenstates of $H_\psi+H_\text{int}$ is complicated, but has been studied extensively in analogous 1D models \cite{vasseur2015particle}, and we may draw lessons from this previous work. Namely, in 1D it was shown, at strong disorder via a real-space renormalization group (RG) treatment, that the interaction terms were preserved under the RG flow, whereas the hopping terms flowed to zero. In the interaction dominated regime, the system naturally breaks the particle-hole symmetry, and forms a particle-hole asymmetric, fully localized state. An essentially identical strong disorder RG-based argument in 2D strongly suggests that the system will flow to the interaction dominated regime, even though the pair-tunneling amplitudes $\Gamma_{ij}$ are typically much larger than the interaction strengths $V_{ij}$ to begin with.

In the present context, the spontaneous particle-hole symmetry broken state corresponds to an MBL phase in which the dynamical $e\leftrightarrow m$ symmetry of the original model is broken, i.e. at strong disorder we expect an MBL Floquet time crystal with eigenstates close (up to finite-depth local unitary transformation) to the form in Eq.~\eqref{eq:tcstate}.

\section{Parafermionic generalizations}
The driven $\Z_2$-topologically ordered example we described  exhibits chiral edge pumping of effectively non-Abelian objects with irrational quantum dimension $\sqrt{2}$, despite that the system's excitations consist only of {\it integer-dimension Abelian anyons}. The resolution to this apparent contradiction is the following: Since we are explicitly driving the system in a time-dependent fashion, energy is not conserved, and we may pump certain confined defects of the Abelian topological order around the edge, without these defects appearing as deconfined bulk quasi-particles. Namely, the Majorana fermions in the above example can be viewed as the ends of topological chains of the emergent fermionic quasi-particles, (or equivalently as ``twist" defects that exchange $e$ and $m$ particles\cite{barkeshli2013twist,barkeshli2014symmetry}), which have irrational quantum dimension $d=\sqrt{2}$. We can readily extend this construction to realize radical CF phases with $\nu = \log \sqrt{N}$ for arbitrary integer $N$, whose edges chirally translate parafermionic defects with quantum dimension $d=\sqrt{N}$.

To this end, we can adapt the driving protocol of Eq.~\eqref{eq:drive}, to the $\Z_N$ generalization of Kitaev's honeycomb model constructed by Barkeshli et al. \cite{barkeshli2015generalized}. We again consider a honeycomb, but replacing the spin-1/2 operators $S_r^i$ with the $N$-state spin operators: $\tau_r^{x,y,z}$, which satisfy $\(\tau_r^i\)^N=1$ and $\tau^z_r = \(\tau^x_r\tau^y_r\)^\dagger$. In addition, the operators on different sites commute, i.e., $[\tau^i_r,\tau^j_{r'}]=0$ for $(r\neq r')$, and on the same site they furnish the algebra
\begin{align}
\tau^x_r\tau^y_r = e^{2\pi i/N}\tau^y_r\tau^x_r,
\end{align}
together with identical relations under the cyclic permutations of the $(x,y,z)$ indices.  Following Ref.~\onlinecite{barkeshli2015generalized}, we may describe the $N$-state spins as quartets of parafermionic twist defects with quantum dimension $\sqrt{N}$ (generalizing the Majorana fermion description for $N=2$), which we can embed spatially around the honeycomb in the $\{c,b^{x,y,z}\}$ positions shown in Fig.~\ref{fig:HoneycombModel}. Generalizing the fusion relations for Majorana defects ($N=2$), these $\Z_N$ twist defects, which we will denote by $\sigma$, can fuse to any number of the anyonic $\psi$ particles: $\sigma\times\sigma = \sum_{j=0}^{N-1}\psi^j$ and $\psi^j \times \sigma = \sigma$. 

As for the spin-1/2 version, it is convenient to pair the bond-centered twist defects, $b^{\ell_{rr'}}$, on bonds $\<rr'\>$ of type $\ell_{rr'}\in \{x,y,z\}$, into $\Z_N$ valued gauge link variables: $\sigma_{rr'} = e^{2\pi ij/N}$, where $j$ is the number of $\psi$ particles in the fusion of the two bond-centered twist defects. Again, the  $\Z_N$ gauge flux through each plaquette:
\begin{align}
F_P = \prod_{\<rr'\>\circlearrowleft P} \tau^{\ell_{rr'}}_{r}\tau^{\ell_{rr'}}_{r'} = \prod_{\<rr'\>\circlearrowleft P} \sigma_{r,r'}
\end{align}
is conserved throughout the Floquet evolution, though this flux operator is now a $\Z_N$ object, having eigenvalues $e^{2\pi i/N}$. Here $\ell_{rr'}$ denotes the type ($x$, $y$, or $z$) of the link $\<rr'\>$. 

On flux-free plaquettes ($F_P=1$), the fusion channel ($\in \{1,\psi,\psi^2,\dots \psi^{N-1}\}$) of the $c$ parafermions on the left and right corners of the hexagon is also conserved. To make contact with the algebraic anyon language describing a  $\Z_N$ gauge theory, we will identify the configuration where there is a plaquette with zero flux, and parafermions fusing to $\psi$, as a $\psi$ particle excitation. Similarly, we will label plaquettes with parafermions fusing to $1$, and a single flux,  ($F_P = e^{2\pi i/N}$) as $m$-excitations, and plaquettes with both parafermions fusing to $\psi$ and $F_P=e^{2\pi i/N}$ as an $e = m\times \psi$ excitation.

Again, the terms $e^{-ih^{[j]}} = \prod_{\<rr'\> \in j}\tau^{j}_r\tau^{j}_{r'}$ exchange the $c$-parafermion defects at the ends of $j$-type bonds. During one Floquet cycle, the two twist defects on the left and right side of a bulk plaquette are braided in a counterclockwise fashion, and hence encircle the $\Z_N$ gauge flux through the plaquette. For plaquettes with $j$ fluxes ($F_P = e^{2\pi i j/N}$), braiding of twist defects originally in the fusion channel $\sigma\times\sigma = \psi^k$ around the flux changes the fusion channel of the twist defects by $\psi^{j-k}$, producing bulk FET order in which the $\Z_N$ gauge charges ($e$) and fluxes ($m$)  are dynamically interchanged.

At the edge, we again see that the Floquet evolution performs a clockwise chiral translation of one parafermionic twist defect per unit cell. Since the parafermion defects have quantum dimension $\sqrt{N}$, this produces an irrational chiral Floquet index: $\nu = \log\sqrt{N}$.

\section{Discussion}
We have so far considered a system with bosonic (spin) degrees of freedom, where the emergence of a radical chiral edge requires Majorana fermion defects arising from emergent fermion degrees of freedom. In fermionic systems where Majorana defects are already present, a radical CF phase with $\nu = \log\sqrt{2}$ can be obtained without any accompanying bulk topological order\cite{po2016chiral,fidkowski2017interacting}. However, for physical problems this either requires the breaking of fermion-number conservation (by pair-superfluidity), which prevents MBL \cite{potter2016symmetry}, or is realized as a pre-thermal phenomenon\cite{else2016pre}.

A natural question to ask is: do these examples exhaust the possible set of Abelian chiral Floquet phases? or do they only represent a partial set? For the systems with $\Z_2$ topological order and $\nu = \pm \log \sqrt{2}$, the edge Floquet evolution either commutes or anticommutes with the bulk evolution depending on the gauge-flux sector of the bulk. The edge of this system can be rigorously characterized by extending the construction of Ref.~\onlinecite{gross2012index} to systems with $\Z_2$ graded tensor product structure~\cite{fidkowski2017interacting}. These results establish that the rational CF phases and radical CF phase with $\nu = \log \sqrt{2}$ form a complete set for systems with $\Z_2$ topological order. This suggests that the radical phases may exhaust the possibilities for other Abelian topological orders, however, rigorously establishing this result would require extending the algebraic construction of \cite{gross2012index,fidkowski2017interacting} to parafermionic algebras, which are comparatively far less well understood, and would require substantial formal mathematical developments, which we leave for future work.

Furthermore, while we have focused on the case of Abelian bulk topological order for the compatibility with MBL\cite{potter2016symmetry}, which can stabilize the system against bulk heating, one could also consider metastable chiral Floquet phases arising in systems with non-Abelian bulk topological order in a prethermal regime\cite{abanin2015exponentially,else2016pre}. A direct anyonic generalization of the bosonic SWAP model of Ref.~\onlinecite{po2016chiral} could be obtained by taking a square lattice of non-Abelian particles, and replacing the SWAP gates by pair-wise braidings, resulting in a chiral translation of non-Abelian anyons at the system boundary. Intuitively, such construction gives rise to a phase with chiral unitary index $\nu = \log d$, where $d$ is the quantum dimension of the anyon in question. Developing a systematic understanding of such non-Abelian CF phases is an important challenge for future work.

We close by briefly commenting on possible experimental signatures of radical CF phases. A crude signature of the chiral edge motion is that it ensures the edge will thermalize regardless of the disorder strength~\cite{po2016chiral}, resulting in decay of non-thermal initial conditions at the edge~\cite{schreiber2015observation}. A more direct signature of the radical chiral edge motion would be to measure the correlation between the state of a spin at site $i$ along the edge at time $t=0$, and at site $i+nT/2$ at time $t=nT$ later. For example, in the honeycomb model, $U(4T)$, $(n=4)$ is precisely the identity in the bulk, and states at the edge will get transferred by $2$ sites along the edge. Finally, the bulk-boundary correspondence ensures that these chiral edge signatures will be accompanied by a bulk time-crystalline order, which can be observed by persistent $2T$-periodic oscillations in generic local observables~\cite{else2016floquet,khemani2015phase,yao2017discrete}.

\textit{Acknowledgements -- }   LF is supported by NSF DMR-1519579 and by Sloan FG-2015- 65244. AV acknowledges support from a Simons Investigator Award and AFOSR MURI grant FA9550-14-1-0035. This research was supported in part by the Kavli Institute of Theoretical Physics and the National Science Foundation under Grant No. NSF PHY11-25915. ACP is supported by NSF DMR-1653007.

\appendix
\section{Computation of the chiral index for the spin-1/2 honeycomb model\label{supp:index}}
In this note, we solve $U(2T)$ exactly by recognizing a connection of the model to the stabilizer formalism of quantum error correction. We will then establish that the chiral unitary index $\nu$ is well-defined despite the bulk topological order, and equal to $\nu[U(2T)] = \log 2$.

The connection to the stabilizer formalism enters since the evolution operator $U(T)$ Eq.~1 takes the special form of a Clifford circuit, whose properties we briefly recount here. Consider a quantum system of $n$ qubits (spin-1/2's) labeled by $r = 1,\dots, n$. For each qubit we have the Pauli operators $X_r$, $Y_r$ and $Z_r$, and we consider the Pauli group of $n$ qubits: $P_n \equiv \{ \alpha \, \Sigma_1 \otimes \Sigma_2 \otimes  \dots \otimes \Sigma_n$ with $\alpha  \in \{1,i,-1,-i\}$, and $\Sigma_r \in \{1,X_r, Y_r, Z_r \} $. We say a unitary operator $U$ is a \textit{Clifford operation} if $U \sigma U^\dagger  \in P_n$ $\forall \Sigma \in P_n$, i.e. if `Pauli products' remain `Pauli products' after conjugation by $U$. Up to an irrelevant overall ${\rm U}(1)$ phase, a Clifford operation is uniquely determined by its action on the Pauli group. One can readily verify that $U(T)$ has this property, which we will exploit to efficiently compute the chiral index of $U(2T)$.

\begin{figure}[h]
\begin{center}
{\includegraphics[width=0.48 \textwidth]{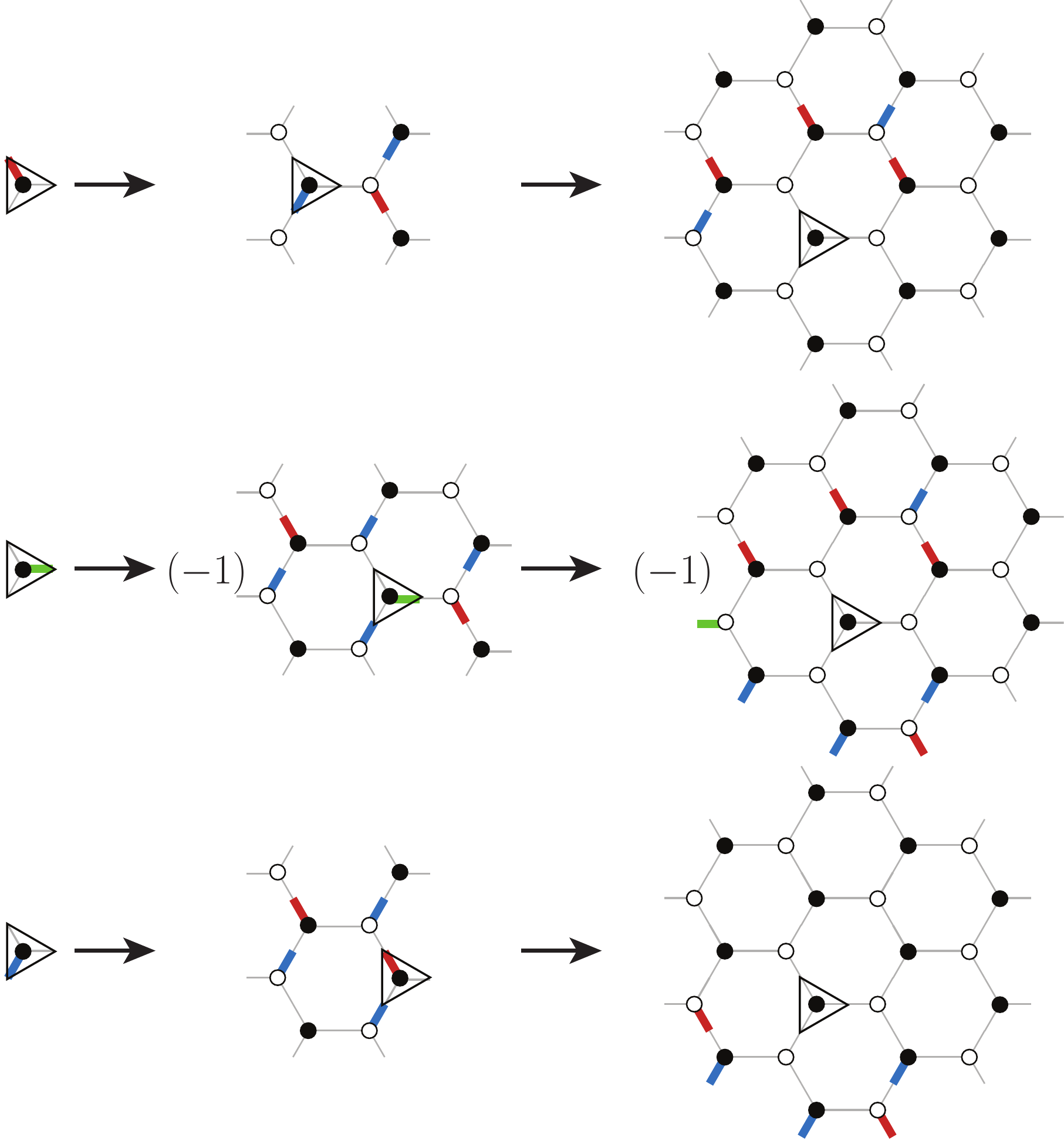}} 
\caption{{\bf Evolution of Pauli operators.} The three Pauli operators associated to each site (filled or open circle) are represented graphically as a colored, thickened line along the corresponding bond. The same color scheme as in Fig.~1 is used, where red, green and blue respectively represent $X_r$, $Y_r$ and $Z_r$.
The triangles indicate the positions of the original sites, and the arrows indicate time evolution by $U(T)$.
\label{fig:Clifford}
 }
\end{center}
\end{figure}

\subsection{Factorization of $U(2T)$ into bulk and edge pieces}
As discussed in the main text, $U(2T)$ admits a complete set of conserved local operators: $\{ \mathcal F_P, \mathcal P_P\}$, where $P$ labels hexagonal plaquettes, which together with the property that $U(2T): c_r\rightarrow \mathcal{F}_{P_r} c_r$ and the translation invariance of the 3-step drive, constrains the form of $U(2T)$ (up to an important overall phase) to:
\begin{equation}\begin{split}\label{eq:UGuess}
U(2T) \overset{?}{=} G_{\phi} \equiv \prod_{P} \left( (1- \mathcal V_P) + e^{i \phi} \mathcal  V_P \mathcal P_P \right),
\end{split}\end{equation}
where $e^{i \phi} $ is a phase, that we will determine shortly.

Since, $U(2T)$ is a Clifford circuit, and $\mathcal{F}_P,\mathcal{V}_P$ are products of Pauli spin operators, $U(2T)$ can only change $\mathcal{F}_P,\mathcal{V}_P$ by phase either $\pm 1$, which requires either $\phi = 0,\pi$, to which we denote the corresponding unitary operators respectively as $G_+$ or $G_-$. One can readily verify that, for periodic $G_\pm$ actually coincide up to an irrelevant factor of $G_+G_-^\dagger = \prod_P \mathcal{F}_P$. For open boundary conditions $G_\pm$ disagree only by a string of Pauli operators at the boundary, which can be implemented by a 1D boundary Hamiltonian, and cannot change the chiral index of the edge. Hence, we are free to consider either $G_\pm$, and for concreteness we will examine $G_+$.

\subsection{Chiral unitary index of $U(2T)$}
Since $U(2T) = G_+$ factorizes into a product of locally commuting terms, its edge is characterized by a rational chiral unitary invariant. From the discussion in Ref.\ \onlinecite{po2016chiral}, with open boundary conditions one can write $U(2T) = Y_{\rm edge} U_{\rm bulk}$ with exponential accuracy, where $Y_{\rm edge}$ is a quasi-1D unitary acting nontrivially only near the edges.
Note that this procedure is unaffected by the fact that $U(2T)$ features intrinsic topological order in the bulk, and so the chiral unitary index of $Y_{\rm edge}$ is well-defined and remains as a diagnostic of the chiral nature of the model.
In addition, the computed index is stable against small perturbation that maintains the MBL nature of the bulk -- and in the present case such robustness can be achieved by appending to the driving protocol a fourth disordering step, as discussed in Sec.\ V of the main text.

To evaluate evaluate $\nu(Y)$, we first recast the original index formula in Ref.\ \onlinecite{gross2012index,po2016chiral} into a form optimized for a Clifford circuit. Recall the overlap $\eta$ of two local operator algebras $\mathcal A$ and $\mathcal B$ is defined as 
\begin{equation}\begin{split}\label{eq:etaDef}
\eta (\mathcal A, \mathcal B) \equiv \frac{\sqrt{p_a p_b}}{p_{\Lambda}} \sqrt{\sum_{\mu=1}^{p_a^2} \sum_{\nu=1}^{p_b^2} \left|\text{Tr}_{\Lambda} \left(  e_{\mu}^{a \dagger}  e_{\nu}^{b \vphantom\dagger} \right) \right|^2},
\end{split}\end{equation}
where $\mu$ ($\nu$) indexes a complete set of basis for $\mathcal A$ ($\mathcal B$). To take advantage of the Clifford structure, we choose a standard basis for an interval with $l$ sites labeled by the multi-index $\mu \equiv (\mu_1,\dots,\mu_l)$, defined through $ \Sigma^L_{\mu} \equiv  \Sigma_{\mu_1}\otimes  \Sigma_{\mu_2}  \otimes \cdots \otimes  \Sigma_{\mu_l}$, where $\mu_i \in \{ 0,1,2,3\}$ labels the Pauli matrices in the standard convention. 

The chiral unitary index is then defined as
\begin{equation}\begin{split}\label{eq:}
\nu(Y) \equiv \log \frac{\eta(Y(\mathcal A_L) , \mathcal A_R)}{\eta( \mathcal A_L, Y(\mathcal A_R) )},
\end{split}\end{equation}
where $\mathcal A_L$ and $\mathcal A_R$ respectively denote the operator algebras (with a sufficiently large size) on the left and right of a specified spatial cut, and $Y(\mathcal A) \equiv \{Ye Y^\dagger ~:~ e\in \mathcal A\}$ is the transformed algebra.

\begin{figure}[h!]
\begin{center}
{\includegraphics[width=0.48 \textwidth]{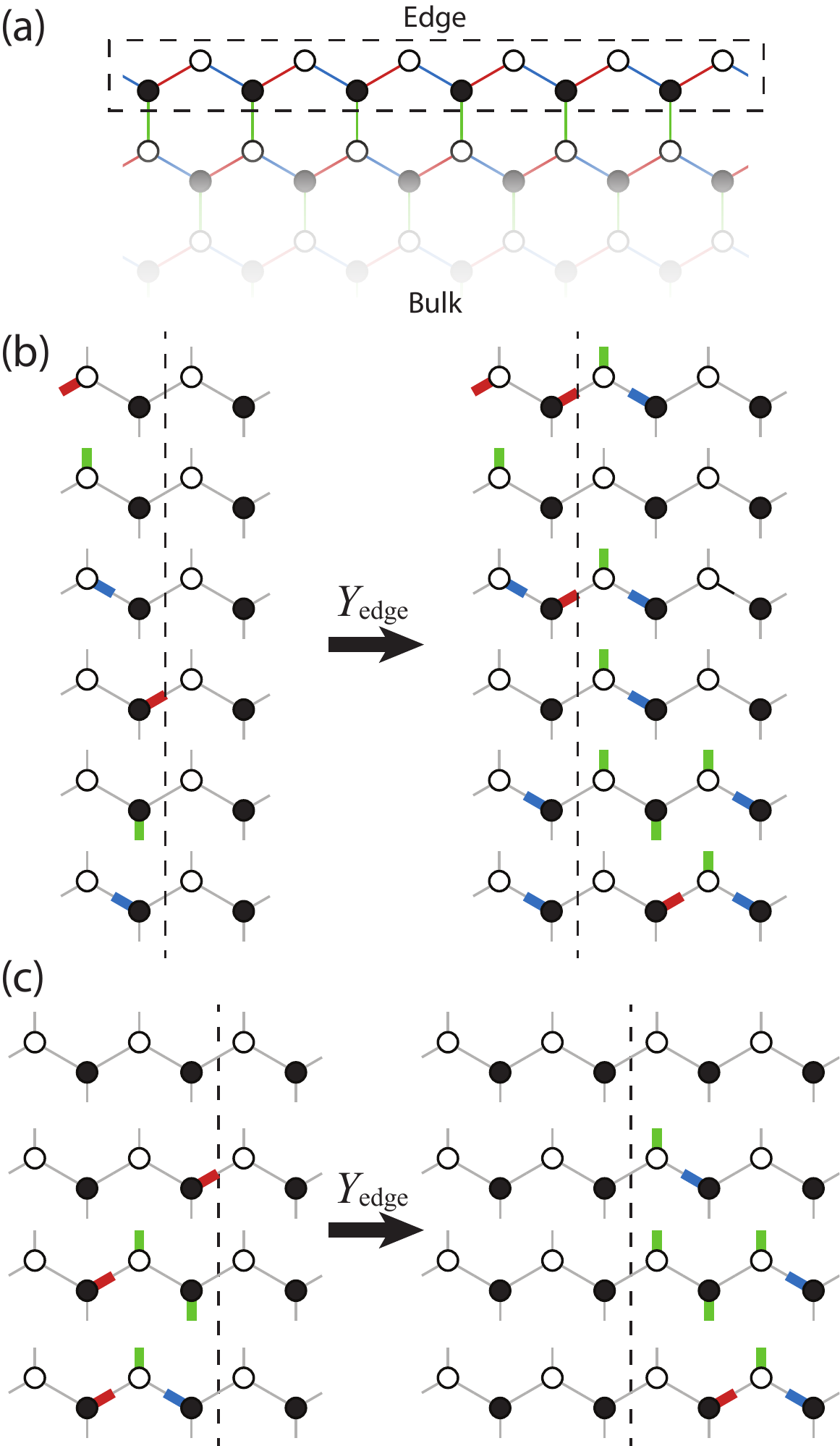}} 
\caption{
{\bf Chiral unitary index $\nu$ of $U(2T)$.} (a) The index is computed for the indicated edge, where the boxed region (of depth being one lattice constant) corresponds to where the edge unitary $Y_{\rm edge}$ acts nontrivially. Note that we have rotated the lattice by $90^\circ$ relative to Fig.~1 of the main text. (b) As the chosen edge retains lattice translation invariance along the parallel direction, $Y_{\rm edge}$ (also a Clifford circuit) is fully specified by computing the evolution of the six Pauli operators associated with the two inequivalent sites in a unit cell. The Pauli operators are represented in the same way as  Fig.~\ref{fig:Clifford}, and we do not keep track of the global phase as it does not enter the index computation. The vertical dashed line indicates a fixed spatial cut.
(c) Using the evolution in (b), one sees that exactly four Pauli operators are `transported' from the left to the right of the cut, and only one (the identity) from right to left. This gives $\nu[U(2T)] = \log \sqrt{4} = \log 2$.
\label{fig:Clifford_GNVW}
 }
\end{center}
\end{figure}

As $\sigma_{\mu_i}$ is traceless for $\mu_i=1,2,3$, only terms with $ \sigma^{a} = \sigma^b$ can contribute in the trace in Eq.~\eqref{eq:etaDef}.
In addition, as $Y$ is a Clifford circuit, generally one finds
\begin{equation}\begin{split}\label{eq:}
Y \left( \bigotimes_{i\in L}  \Sigma_{\mu_i} \right)  Y^\dagger= \pm   \left( \bigotimes_{i\in L}  \Sigma_{\mu_i'} \right)  \otimes \left( \bigotimes_{j\in R}   \Sigma_{\nu_j} \right),
\end{split}\end{equation}
and we say $ \bigotimes_{i\in L}  \Sigma_{\mu_i} $ is `transported across the cut' if $\mu_i' = 0 ~\forall i$, i.e.~$\left( Y     \Sigma_{\mu}^L  Y^\dagger \right)|_{L} =  1$ (this includes, in particular, the identity). These are the only operators that can contribute in $\eta( Y  \mathcal A_L Y^\dagger, \mathcal A_R)$, and each such term contribute with the same weight as the identity. Therefore the index formula for a Clifford circuit is simply a counting formula:
\begin{equation}\begin{split}\label{eq:}
\nu(Y) \overset{\text{Clifford}}{=} \log \sqrt{\frac{| \{  \Sigma^L_{\mu}~:~ \left( Y     \Sigma_{\mu}^L  Y^\dagger \right)|_L =  1 \} | }{| \{  \Sigma^R_{\nu}~:~ \left( Y     \Sigma_{\mu}^R  Y^\dagger \right)|_R =  1 \} |}}.
\end{split}\end{equation}

The computation is detailed in  Fig.~\ref{fig:Clifford_GNVW}, which shows that $\nu[U(2T)] = \log 2$ -- the minimal rational value.
As discussed in the main text, this implies $\nu[U(T)] = \frac 12 \nu[U(2T)] = \log \sqrt{2}$, which falls outside of the original GNVW classification. Such a radical index is allowed because, unlike $U(2T)$, $U(T)$ cannot be factorized into commuting bulk and edge pieces due to the bulk FET order.

\bibliography{FloqSPTbib}

\end{document}